\def\half{\mbox{$\frac{1}{2}$}}         
\def\b{\overline}
\def\s{\varsigma}

\documentstyle[12pt]{article}
\makeatletter
\def\setb@se#1{\baselineskip=#1 \normalbaselineskip=#1}
\setb@se{12pt}
\long\def\title#1{\vspace*{11.5pc}{\pretolerance=10000\raggedright
  \setb@se{12pt}\bf #1\par}\nobreak\ignorespaces}
\long\def\author#1{\vspace{4pc}\begin{list}{\hfill}%
{\topsep=0pt\parskip=0pt\parsep=0pt\partopsep=0pt\listparindent=0pt%
\itemsep=0pt\rightmargin=0pt\labelsep=0pt\labelwidth=5pc\leftmargin=5pc}%
\item\normalsize{#1}\end{list}\vspace{14pt}}
\long\def\affil#1{\begin{list}{\hfill}%
{\topsep=0pt\parskip=0pt\parsep=0pt\partopsep=0pt\listparindent=0pt%
\itemsep=0pt\rightmargin=0pt\labelsep=0pt\labelwidth=5pc\leftmargin=5pc}%
  \item\normalsize{\rm #1}\end{list}\vspace{7pt}}
\long\def\beginabstract{\vspace{21pt plus 7pt minus 7pt}\begin{list}{\hfill}%
{\topsep=0pt\parskip=0pt\parsep=0pt\partopsep=0pt\listparindent=0pt%
\itemsep=0pt\rightmargin=0pt\labelsep=0pt\labelwidth=5pc\leftmargin=5pc}%
\item\normalsize{\bf Abstract. }}
\long\def\endabstract{\end{list}\vspace{28pt plus14pt minus 14pt}%
\normalsize\noindent}
\def\@sect#1#2#3#4#5#6[#7]#8{\ifnum #2>\c@secnumdepth
  \def\@svsec{}\else
  \refstepcounter{#1}\edef\@svsec{\csname the#1\endcsname.\hskip 0.5em}\fi
  \@tempskipa #5\relax
   \ifdim \@tempskipa>\z@
     \begingroup #6\relax
       \@hangfrom{\hskip #3\relax\@svsec}{\interlinepenalty \@M #8\par}%
     \endgroup
    \csname #1mark\endcsname{#7}\addcontentsline
      {toc}{#1}{\ifnum #2>\c@secnumdepth \else
                   \protect\numberline{\csname the#1\endcsname}\fi
                 #7}\else
     \def\@svsechd{#6\hskip #3\@svsec #8\csname #1mark\endcsname
                   {#7}\addcontentsline
                        {toc}{#1}{\ifnum #2>\c@secnumdepth \else
                          \protect\numberline{\csname the#1\endcsname}\fi
                    #7}}\fi
  \@xsect{#5}}

\def\section{\@startsection {section}{1}{\z@}{-28pt plus -14pt minus
-14pt}{2.3ex plus .2ex}{\normalsize\bf}}
\def\subsection{\@startsection{subsection}{2}{\z@}{-14pt plus -8pt
minus -4pt}{1.5ex plus .2ex}{\normalsize\bf}}
\def\subsubsection{%
    \@startsection{subsubsection}{3}{\z@}{-14pt plus
    -8pt minus -4pt}{-1.5ex plus -.2ex}{\normalsize\bf}}
\def\paragraph{\@startsection
     {paragraph}{4}{\z@}{3.25ex plus 1ex minus .2ex}{-1em}{\normalsize\bf}}
\def\subparagraph{\@startsection
     {subparagraph}{4}{\parindent}{3.25ex plus 1ex minus
     .2ex}{-1em}{\normalsize\bf}}

\def\caption{\refstepcounter\@captype \@dblarg{\@caption\@captype}}
\long\def\@caption#1[#2]#3{\par\addcontentsline{\csname
 ext@#1\endcsname}{#1}{\protect\numberline{\csname
 the#1\endcsname}{\ignorespaces #2}}\begingroup
   \@parboxrestore
   \hspace*{28pt}
   \parbox{394pt}{\@makecaption{{\bf\csname
            fnum@#1\endcsname}}{\ignorespaces #3}}\par
\vspace{7pt}\endgroup}

\long\def\@makecaption#1#2{
   \vskip 14pt
   \setbox\@tempboxa\hbox{#1{\bf.} #2}
   \ifdim \wd\@tempboxa >\hsize   
       #1{\bf.} #2\par            
     \else                        
       \hbox to\hsize{\box\@tempboxa\hfil}
   \fi}

\newcommand{\boldarrayrulewidth}{1pt} 

\def\bhline{\noalign{\ifnum0=`}\fi\hrule \@height
                    \boldarrayrulewidth \futurelet
                    \@tempa\@xhline}

\def\@xhline{\ifx\@tempa\hline\vskip \doublerulesep\fi
      \ifnum0=`{\fi}}

\def\thebibliography#1{\section*{REFERENCES\@mkboth
  {REFERENCES}{REFERENCES}}\list
  {\hfil[\arabic{enumi}]}{\itemsep=0pt\labelsep=7pt\itemindent=-14pt
    \settowidth\labelwidth{[#1]}
    \leftmargin\labelwidth
    \advance\leftmargin\labelsep
    \advance\leftmargin -\itemindent
    \usecounter{enumi}}\setb@se{12pt}\small
    \def\newblock{\hskip .11em plus .33em minus .07em}
    \sloppy\clubpenalty4000\widowpenalty4000
    \sfcode`\.=1000\relax}
\def\references{\section*{REFERENCES\@mkboth
{REFERENCES}{REFERENCES}}\list{}{\itemsep=0pt\labelsep=0pt\itemindent=-28pt
\labelwidth=0pt\leftmargin=28pt}\setb@se{12pt}\small
\def\newblock{\hskip .11em plus .33em minus .07em}
\sloppy\clubpenalty4000\widowpenalty4000
\sfcode`\.=1000\relax}
\newcommand{\dash}{------}

\newcommand{\APJ}{{\em Astrophys.\ J.} }

\newcommand{\JMP}{{\em J. Math.\ Phys.} }

\newcommand{\PR}{{\em Phys.\ Rev.} }
\newcommand{\PRL}{{\em Phys.\ Rev.\ Lett.} }

\makeatother

\begin{document}


\title{THE MIXMASTER COSMOLOGICAL METRICS\footnote{to appear
                in D. Hobill, ed., {\em Deterministic Chaos
                in General Relativity}, Plenum Pub.\ Co.\ 1994}
                \footnote{listed as gr-qc/9405068}}

\author{Charles W. Misner \footnote{e-mail: Misner@umdHEP.umd.edu}}

\affil{Dept.\ of Physics,
University of Maryland, College Park MD 20742-4111}

\beginabstract
This paper begins with a short presentation of the Bianchi IX or
``Mixmaster'' cosmological model, and some ways of writing the
Einstein equations for it.  There is then an interlude describing how
I came to a study of this model, and then a report of some mostly
unpublished work from a Ph.\ D. thesis of D. M. (Prakash) Chitre
relating approximate solutions to geodesic flows on finite volume
negative curvature Riemannian manifolds, for which he could quote
results on ergodicity.  A final section restates studies of a
zero measure set of solutions which in first approximation appear to
have only a finite number of Kasner epochs before reaching the
singularity.  One finds no plausible case for such behavior in better
approximations.
\endabstract

\section{BIANCHI {\sf IX} COSMOLOGY}

\indent
The ``Mixmaster'' universe is an empty homogeneous cosmology of
Bianchi type IX, i.e., it a (class of) solutions of Einstein's vacuum
equations with three dimensional spacelike surfaces which are orbits
of a particular symmetry group. The spacetime manifold $M$ is $R
\times S^3 = R^4 - \{ 0 \}$ with the differential structure inherited
from the standard euclidean coordinates $wxyz$ of $R^4$. When $R^4$
is presented as quaternions, $M$ is the manifold of all $q \neq 0$
where $q = w + ix + jy + kz$.

The symmetry group is the simply connected covering group of SO(3)
and is realized as the unit quaternions $|p| = 1$ acting as
transformations $q \mapsto pq$.  These mappings of $M$ onto
itself preserve $|q|^2 = \b{q}q = w^2 + x^2 + y^2 + z^2$ and leave
invariant the differential forms
\begin{eqnarray}
\sigma_x & = & \frac{1}{\overline{q}q}[i\b{q}\,dq + \b{(i\b{q}\,dq)}]
                \\[\medskipamount]
	 & = & \frac{2}{|q|^2}[x\,dw - w\,dx - z\,dy + y\,dz]
           \nonumber
\end{eqnarray}
and $\sigma_y, \sigma_z, dt$ which are defined by similar formulae
where $i$ above is replaced by $j, k, 1$ respectively.  [Note that
quaternion conjugation satisfies\ $\b{(pq)} = \b{q}\,\b{p}$ and that
$ij = -ji = k$ etc.]
{}From these definitions one can verify $d\sigma_x = \sigma_y \wedge
\sigma_z$ and cyclic permutations thereof in which one sees the
structure constants of $SO(3)$.  One also integrates $dt$ to find
that $|q|^2 = e^t$.

The metric on $M$ is taken to be
\begin{equation}
ds^2 = - N^2(t)dt^2 + g_{ij}(t)\,\sigma_i \sigma_j \quad .
\end{equation}
It will be invariant under the transformations $q \rightarrow pq$
with $|p| = 1$ since $t$ and the $\sigma_i$ are. Then the
coefficients are parameterized as
\begin{eqnarray}
g_{xx} & = & \exp[2(-\Omega + \beta_+ + \sqrt{3}\beta_-)] \nonumber\\
g_{yy} & = & \exp[2(-\Omega + \beta_+ - \sqrt{3}\beta_-)]
                            \label{e-gij}\\
g_{zz} & = & \exp[2(-\Omega - 2\beta_+)]                  \nonumber
\end{eqnarray}
with off-diagonal $g_{ij} = 0$.  This lets $\Omega$ control the
3-volume $\sqrt{{}^3g} = e^{-3\Omega}$ while the $\beta$'s control the
shape or shear deformation of the 3-geometry.

\subsection{ADM Hamiltonian}

\indent
The Einstein equations can be summarized in a variational
principle $\delta I = 0$ with
\begin{equation}
I = \int (p_+\,d\beta_+ + p_-\,d\beta_- - H\,d\Omega)
\end{equation}
where
\begin{equation}
H^2 = p_+^2 + p_-^2 + e^{-4\Omega}(V - 1)
\end{equation}
and $V = V(\beta_+, \beta_-)$.  To reconstruct the metric from
solutions of Hamilton's equations one must select a time coordinate.
The choice $t = \Omega$ requires that one set $N \propto
H^{-1}e^{-3\Omega}$ which can lead to difficulties when $H=0$ but is
not problematic near the singularity $\Omega \rightarrow \infty$.
[For careful treatment of the scale factors for
coordinates and the action $I$, see Misner (1969b).]
The function $V(\beta)$ which appears here is called the
anisotropy potential and is defined by
\begin{eqnarray}
  V & = & \mbox{$\frac{1}{3}$} e^{-8\beta_+}
	   - \mbox{$\frac{4}{3}$} e^{-2\beta_+}\cosh 2\sqrt{3}\beta_-
                    \nonumber\\
    &   &  + 1
      +\mbox{$\frac{2}{3}$}e^{4\beta_+}(\cosh 4\sqrt{3}\beta_- - 1)
\end{eqnarray}
This function is positive definite and has symmetry under $2\pi/3$
rotations in the $\beta_+ \beta_-$ plane.
It can also be written
\begin{equation}\label{e-Vtr}
  V = \mbox{$\frac{1}{3}$}\mbox{\rm tr} (e^{4\beta} - 2 e^{-2\beta} +1)
\end{equation}
where $g_{ij} = e^{-2\Omega}(e^{2\beta})_{ij}$ for a symmetric
traceless matrix $\beta$.
The most important asymptotic form is
\begin{equation}\label{e-wall}
  V \sim  \mbox{$\frac{1}{3}$} e^{-8\beta_+} , \quad |\beta_-| <
            -\sqrt{3} \beta_+
\end{equation}
which defines the typical potential wall, and
\begin{equation}
  V - 1 \sim  16e^{4\beta_+}\beta_-^2 - \mbox{$\frac{4}{3}$} e^{-2\beta_+},
                \quad |\beta_-|\ll 1,\; \beta_+ > 1
\end{equation}
which shows a ``corner'' opposite the typical potential wall.

\subsection{Constrained Hamiltonian}

\indent
A nicer hamiltonian is obtained by taking $\Omega$ as a dynamical
variable and introducing $\tau$ through $d\tau = d\Omega/H$
as the independent path parameter.  Then there are three degrees of
freedom and the new hamiltonian is
\begin{equation}\label{e-H}
 {\cal H} = \half[-p_\Omega^2 + p_+^2 + p_-^2 + e^{-4\Omega}(V - 1)]
\end{equation}
but the constraint ${\cal H}=0$ must be appended to Hamilton's
equations to reproduce the full content of Einstein's equations.  The
lapse function when $\tau$ is used as a coordinate time is then $N =
e^{-3\Omega}$.  [See (Misner 1972) where the coordinate here called
$\tau$ was called $\lambda$. The present notation agrees with Hobill
(1994) from BKL apart possibly from some numerical factor.]

Some simple motions can be read from this form.
At fixed $\beta$ the $e^{-4\Omega}(V - 1)$ term becomes negligible as
$\Omega$ increases.  Neglecting this term in ${\cal H}$ makes each
$p$ constant with $p_\Omega^2 = p_+^2 + p_-^2$. These are the Kasner
epochs where $(d\beta_+/d\Omega)^2 + (d\beta_-/d\Omega)^2 = 1$ since
Hamilton's equations give $d\beta_\pm/d\Omega = - p_\pm/p_\Omega$.
The potential term in equation~(\ref{e-H}) is essentially the scalar
curvature of the 3-space ${}^3R = - 6 e^{2\Omega}(V - 1)$ so ignoring
it gives the dynamics of the spatially flat Bianchi I solutions.  To
connect this notation with common parameterizations of the Kasner
solutions (Lifshitz and Khalatnikov 1963, Hobill 1994) we must set
\begin{equation}\label{e-BKL}
\begin{array}{rcrcr}
  p_1  & = & (p_+ + \sqrt{3}p_- + p_\Omega)/3p_\Omega & = &
-u/(u^2+u+1) \\
  p_2  & = & (p_+ - \sqrt{3}p_- + p_\Omega)/3p_\Omega & = &
(1+u)/(u^2+u+1) \\
  p_3  & = & (-2p_+ + p_\Omega)/3p_\Omega & = &
u(1+u)/(u^2+u+1)
\end{array}
\end{equation}
If one chooses evolution toward the singularity so $d\Omega/d\tau >
0$ then when these formulae are valid $p_\Omega = -\sqrt{p_+^2 +
p_-^2}$.

{}From the Hamilton equations $dp_\Omega/d\tau =
-\partial{\cal H}/\partial\Omega$ and $d\Omega/d\tau = - p_\Omega$
one derives
\begin{equation}
  \frac{d}{d\tau}\left(p_\Omega^2 + e^{-4\Omega}\right) = 4p_\Omega
        e^{-4\Omega} V
\end{equation}
which from $V \geq 0$ shows that approaching the singularity
($p_\Omega < 0$) the positive quantity $(p_\Omega^2 + e^{-4\Omega})$
always decreases but remains constant except when the system point is
interacting with a potential wall.  Near the singularity
($e^{-4\Omega}\rightarrow 0 $) one expects these properties to hold
for $p_\Omega^2$. Rigorous control on the behavior of $p_\Omega$ near
the singularity would be very useful.  In Misner (1970) it is
suggested from a quasiperiodic solution that $p_\Omega \Omega$ has
only a limited percentage variation.  Also easily verified from
Hamilton's equation and ${\cal H} = 0$ is
\begin{equation}
  \frac{d}{d\tau}\left(p_\Omega  e^{2\Omega}\right) =- 2
        e^{2\Omega} (p_+^2 + p_-^2) \leq 0
\end{equation}
which in these variables gives the proof (Rugh 1990, Rugh and Jones
1990) that once the volume of the universe $e^{-3\Omega}$ begins to
decrease it can never stop decreasing.  For once $-d\Omega/d\tau =
p_\Omega$ is negative (volume decrease) the quantity $p_\Omega
e^{2\Omega}$ must never increase, so $p_\Omega$ can never thereafter
become zero.

In a bounce against a single potential wall described by
equation~(\ref{e-wall}) there are two constants of motion, $p_-$ and
$K \equiv \half p_+ - p_\Omega$ from which the change in $u$ and in
$p_\Omega$ can be deduced (Misner 1969b).
The location where the potential wall becomes significant is fixed by
$p_\Omega^2 e^{4\Omega} = (V - 1)$ which gives
\begin{equation}
  \beta^w_+ = -\half\Omega - \mbox{$\frac{1}{8}$} \ln (3p_\Omega^2)
            \quad .
\end{equation}

\section{EARLY MIXMASTER}
\indent
My involvement with the Mixmaster cosmologies did not begin with the
horizon problem, but with questions of anisotropy in the recent
universe.  Beginning a sabbatical year in Cambridge, I resumed
keeping notebooks following the example of my mentor John Wheeler.
One notation (16 Nov '66) indicates the speculations that provoked my
attention to anisotropic cosmology.
\begin{quote}
Last night Faulkner and Strittmatter showed me their
blackboard globe \ldots\ showing that the $z>1.5$ quasars (all 13 of
them) were all \ldots\  near the galactic poles.  Although this
appears most likely either an ununderstood observational bias, or
\ldots\, we felt that calculations showing its (probably extreme)
implications if interpreted in terms of anisotropic cosmologies,
should be carried out, and today set about this.
\end{quote}
I initially studied Bianchi Type I examples as later described in
(Misner 1968). But my previous acquaintance with Bianchi Type IX in
work with Taub (Misner 1963, Misner and Taub 1968) led me to extend
these techniques to that case.  A notebook entry in the spring (20
Mar '67) begins
\begin{quote}
This is another curvature computation showing that gravitational
fields can, like collisionless radiation, give rise to anisotropy
potential energy.
\end{quote}
and the next day I recorded a definition of the anisotropy potential
of equation~(\ref{e-Vtr}):
\begin{quote}
Let us study this anisotropy potential. Define
$$
V_g = (1/4) \sum_k \left( e^{4\beta_k} - 2 e^{-2\beta_k} + 1
                   \right)
$$
\ldots
\end{quote}
These studies were not published accessibly at that time, but were
included in an April 1967 essay that won the third prize in the
Babson Gravity Research Foundation Essay competition that spring.
Although the analysis included the idea of $\beta$ colliding with the
potential walls, its emphasis was on the decrease of anisotropy
during expansion as related to the cosmic microwave radiation and on
the refocussing of cosmological theory from measuring the FRW
constants to explaining why we live in an FRW universe:
\begin{quote}
The most accurate observations of this radiation concern
its isotropy \ldots\ surely
deserves a better explanation \ldots\

The traditional problem has been to \ldots\ distinguish among the
small number of homogeneous and isotropic cosmological solutions
\ldots\ .

\ \ldots\ we propose \ldots\ wider problem \ldots\ explanation of
whatever degree of homogeneity and isotropy the observations \ldots\
reveal.
\end{quote}
These concerns were repeated in (Misner 1968) where to simplify the
presentation the analysis was limited to Bianchi type I.  The
Mixmaster was published later (Misner 1969a) when I had conjectured
that it would contribute to the horizon problem.  Several months
before this Belinskii and Khalatnikov (1969) had completed work on an
independent analysis of these same solutions which appeared in
Russian within a month of my Phys.\ Rev.\ Letter.  Their work was
based on earlier unpublished analyses by E. M. Lifshitz and I. M.
Khalatnikov in 1962 on a simpler Bianchi type with only two potential
walls which introduced the important description $u \mapsto u-1$ for
change from one Kasner epoch to another and the $u \mapsto 1/u$
change marking the end of this sequence/era.  The aim of these Russian
works was to elucidate the character or existence of a physical
singularity in the generic solution of Einstein's equations.

\section{TOWARD GEODESIC FLOW}

\indent
Let us now return to a Hamiltonian description of the dynamics with
the aim of emphasizing the repetitive aspects of the evolution toward
the singularity.  One can nearly halt the motion of the potential wall
by a new metric parameterization:
\begin{eqnarray}
    \beta_+ & = & e^t \sinh\zeta \cos\phi \nonumber\\
    \beta_- & = & e^t \sinh\zeta \sin\phi \label{e-tzeta}\\
    \Omega  & = & e^t \cosh\zeta          \nonumber
\end{eqnarray}
together with a modified independent variable $d\lambda =
e^{-2t}\,d\tau$ which rescales the Hamiltonian so ${\cal H}e^{2t} =
\tilde{{\cal H}}$. The result is a Hamiltonian system based on
\begin{equation}\label{e-H'}
  2\tilde{{\cal H}} = -
  p_t^2 + p_\zeta^2 + (p_\phi / \sinh\zeta)^2
			    + e^{2t}e^{-4\Omega}(V-1) \quad .
\end{equation}
This Hamiltonian is conserved since it is independent of $\lambda$
and must be initialized as $\tilde{{\cal H}} = 0$ to yield
Einstein's equations.
The potential wall's location is now given by $p_t^2 =
e^{2t}e^{-4\Omega}(V-1)$ which using~(\ref{e-wall}) and
(\ref{e-tzeta}) becomes
\begin{equation}\label{e-wall'}
\sinh\zeta\cos\phi +
\half \cosh\zeta = \mbox{$\frac{1}{8}$} e^{-t}[2t - \ln (3p_t^2)]
\end{equation}
which for large $t$ becomes independent of $t$.  Note that when the
$(V-1)$ term in (\ref{e-H'}) can be neglected $p_t = \Omega p_\Omega
+ \beta_+ p_+ + \beta_- p_-$ is constant.  But also $p_t$ is constant
when this term can be considered independent of $t$, which occurs for
a very steep stationary potential wall, i.e., for large $t$ when a
steepness factor $e^t$ appears in the exponents of
$\exp(-4\Omega-8\beta_+) = \exp[-4e^t(\cosh\zeta + 2\sinh\zeta
\cos\phi)]$.  In this case the potential term is effectively either
zero or infinite crossing a fixed $\zeta(\phi)$ curve and thus without
effective $t$ dependence.  A rigorous limit on the change possible in
$p_t$ would be desirable, but I will work under the expectation that
$p_t$ is constant for $t \rightarrow \infty$.  Then since Hamilton's
equations from~(\ref{e-H'}) give $dt/d\lambda = \partial\tilde{{\cal
H}}/\partial p_t = - p_t$ we see that asymptotically $t$ is
just $\lambda$ rescaled by $-p_t > 0$.

In his Ph.\ D. thesis D. M. Chitre (1972a) recognized that the metric
$d\s^2 = d\zeta^2 + \sinh^2\zeta\,d\phi^2$ whose inverse appears in
equation~(\ref{e-H'}) was that of the Lobachevsky plane so that
results from classical hyperbolic geometry could be employed.  In
particular, when the potential term in equation~(\ref{e-H'}) can be
ignored the solutions will be geodesics on the hyperbolic plane.  He
employed a useful standard coordinate system on this plane by setting
$\sinh\zeta = 2r/(1-r^2)$ to find $d\s^2 = 4(dr^2 + r^2
d\phi^2)/(1-r^2)^2$ so a complex coordinate $z = x + iy$ could be
introduced giving a metric for the (geodesic) motion of the system
point
\begin{equation}
  d\s^2 = \frac{4 dz\,d\b{z}}{(1-|z|^2)^2}
\end{equation}
which is conformally flat. But this metric is invariant under the
transformations
\begin{equation}\label{e-xfn}
z \mapsto \frac{az+b}{\b{a} + \b{b}z}
        {\textstyle \rm \ \; with\ \;} |b| < |a|
\end{equation}
so these transformations which map the unit disc onto itself are
conformal transformations.  They also map circles into circles
and the unit circle onto itself, with straight lines in the complex
plane counted as circles.  Using these transformations any
geodesic of $d\s^2$ can be mapped onto the imaginary axis.  Thus the
geodesics are the arcs of circles in the complex plane that meet the
boundary circle $|z| = 1$ perpendicularly.  In these coordinates the
location of the potential wall at negative $\beta_+$ is given by
translating equation~(\ref{e-wall'}) to find (with $\phi = \pi$ to
locate the wall's intersection $x_w$ with the real axis)
\begin{equation}\label{e-xw}
  x_w = \frac{-2+\sqrt{3-f^2}}{1+f}
 \end{equation}
where
\begin{equation}
  f = \mbox{$\frac{1}{4}$} e^{-t}[2t - \ln (3p_t^2)] \quad .
\end{equation}
The connection between the parameters $(t,x,y)$ and the metric has by
now gotten rather exotic:
\begin{equation}
   ds^2 = -e^{4t}\exp(-6e^t\frac{1+|z|^2}{1-|z|^2})\,d\lambda^2
                  + g_{ij}(t,z)\sigma_i \sigma_j
\end{equation}
where $g_{ij}(t,z)$ is defined through equations~(\ref{e-gij}),
(\ref{e-tzeta}) and $e^{i\phi}\sinh\zeta = 2z/(1-|z|^2)$.  Note that
the $tz$ and $t\zeta\phi$ parameterizations of the metric include
only a sector $\Omega > \sqrt{\beta_+^2 + \beta_-^2}$ of the complete
minisuperspace of kinematically possible 3-geometries.  This sector
includes all regions known to be dynamically accessible as the
singularity $t \rightarrow \infty$ is being approached, but it does
not include regions of negative $\Omega$ where a maximum of volume
expansion can occur.  Geodesics $z(\lambda)$ in the Lobachevsky plane
can be lifted by $t = -p_t \lambda$ to geodesics in minisuperspace
with $d\$^2 = - dt^2 + 4 dz\,d\b{z}(1-|z|^2)^{-2}$ where they represent
Kasner evolution of the 3-geometry.

But not only the potential-free (Kasner) motion can be described by
geodesics in minisuperspace.  The bounce from one Kasner
epoch to the next can also be handled by these techniques.  The
potential wall can be idealized as steep and stationary by setting
$f=0$ in equation~(\ref{e-xw}) so $p_t$ is constant even during the
bounces.  But also, the potential walls are themselves geodesics as
drawn in Figure~1.
The one described by equation~(\ref{e-wall'}) in the stationary
approximation is just the circle $|z+2|^2 = 3$.  By the
transformation~(\ref{e-xfn}) with $a=1$ and $b = 2-\sqrt{3}$ Chitre
makes this circle become the imaginary axis.  The Hamiltonian is now
\begin{equation}\label{e-Hxy}
  2\tilde{{\cal H}} = - p_t^2
  + \mbox{$\frac{1}{4}$}(p_x^2 + p_y^2)(1-x^2-y^2)^2
			    + 2\tilde{{\cal V}}(t,x,y) \quad .
\end{equation}
The approximation of a fixed infinitely steep potential wall makes
$\tilde{{\cal V}}$ independent of $t$, and a transformation has set
the wall of interest to the $y$-axis, so $\tilde{{\cal V}} = \tilde{{\cal
V}}(x)$.  Thus the bounce is just a specular reflection with
$p_t$ and $p_y$ constant while $p_x$ changes sign.

Next Chitre plays the hall of mirrors game.  The potential well has
become a triangle made up of three equivalent steep walls located on
geodesics that are circles in the $xy$-plane.
\begin{figure}[tb]
\vspace*{2in}
\caption[Potential Walls in the Lobachevsky Plane.]{
Potential Walls in the Lobachevsky Plane.  The shape of a spatial
slice of the Mixmaster cosmology is described by the values of a
point $z = x+iy$ within the unit disk in the complex plane.  The
dynamic evolution of this shape is controlled by the
Hamiltonian~(\protect\ref{e-Hxy}). In the sketch of the unit disk
above on the left, the right hand geodesic triangle represents the
location of the potential walls (at large $t$, i.e., near the
singularity). The smaller geodesic triangle is a mirror image.  An
isometric transformation~(\protect\ref{e-xfn}) taking the common side
into a diameter of the unit disk would be needed to make their
congruence manifest.  In the sketch of the unit disk above on the
right further mirror images of these triangles in their sides are
shown. Also shown is a geodesic generating the evolution of the
universe.  When its segments in other images are mapped back into
equivalent arcs in the fundamental triangle (potential well), it
describes the evolution of the shape of space approaching the
singularity.}
\end{figure}
One wall has been translated into the $y$-axis.  Instead of watching
the system point interrupt its geodesic motion by bouncing against
this wall, we can equivalently let the geodesic continue but, by an
$x \mapsto -x$, $y \mapsto y$ transformation, move the potential well
to the other side of the $y$-axis.  Instead of seeing the system
point bounce we watch its mirror image continue the geodesic on the
other side of the mirror where it will be headed toward another
potential wall in the reflected potential.  There it can again
continue into another reflection image of the potential well.  Since
the hyperbolic plane is homogeneous and isotropic, every bounce and
every reflection is equivalent under some symmetry to the first one
analysed.

The limiting (steep and static) potential well defines a fundamental
domain in the Lobachevsky plane which is a triangle bounded by three
equivalent geodesics that intersect at infinity ($|z| = 1$).  This
domain is not compact but has finite area in the $d\s^2$ metric. By
isometries that give reflections in these walls, one obtains a tiling
of the hyperbolic plane and a single geodesic running through an
infinity of tiles can be projected back to a succession of bounces in
the fundamental domain. Using results from Hopf (1936) and Hedlund
(1939) for geodesic flows on quotients of the Lobachevsky plane by a
Fuchsian group, Chitre verifies that the flow is ergodic by verifying
the hypotheses of theorems showing that the geodesic flow here is
metrically transitive and also mixing.  To define these term we need
to introduce a measure on the initial conditions or tangent
directions for these geodesics.
(One assumes $p_t = -1$ in this $t \rightarrow
\infty$ limit where the $t$ dependence has by assumption dropped out
of the problem.)
The tangent directions $(x,y,\theta)$ are a point $(x,y)$
in the fundamental domain and a direction $\theta$ for the unit
tangent vector there.  The measure chosen is defined by the integral
\begin{equation}\label{e-measure}
  \int\int\int\frac{4dx\,dy\,d\theta}{(1-x^2-y^2)^2}
\end{equation}
and is invariant under the transformations~(\ref{e-xfn}) extended to
the circle bundle of unit vectors tangent to this manifold.  In terms
of this invariant measure metrically transitive means that given any
two positive measure measurable sets $M$ and $N$ of tangent
directions (initial conditions for the Mixmaster universe near the
singularity) the geodesics defined by $M$ after some length $\lambda$
(evolving toward the singularity) will define tangent directions
$M_\lambda$ of which some coincide with tangent directions in $N$.
Equivalently, metric transitivity means that every measurable set of
tangent directions that is invariant under the geodesic flow has
either measure zero or full measure (the measure of the whole bundle
of tangent directions over the fundamental domain).  Mixing means
that as a set $M$ of positive measure is evolved along an interval of
$\lambda$ to the set $M_\lambda$ it will asymptotically ($\lambda
\rightarrow \infty$) overlap any measurable set of tangent directions
in exactly the ratio of $M$'s measure that of the whole space (which
is $2\pi$ times the area of the fundamental domain).  Thus after a
long interval of $\lambda = t$ any positive measure set of
$(x,y,\theta)$ tangent directions gets spread out uniformly (although
perhaps thinly if the initial set was small) among all such tangent
directions.
Chitre's summary is
\begin{quote}
This then characterizes the `chaos' of the universe.  If we start
with a well-defined state of the universe and let it evolve towards
the singularity, we find that it goes through almost all possible
anisotropic stages.
\end{quote}

\section{EVALUATING GEODESIC FLOW}

\indent
In interpreting the geodesic flows on the Lobachevsky plane, one must
be aware that as a plausible approximation to solutions of the
Einstein equations only one direction $t \rightarrow +\infty$ on the
geodesic can be accepted.  The opposite direction toward small $t$
leads to conditions where the motion of the potential wall cannot be
neglected.  But there are geodesics where even the proper
direction leads to problematic situations.  Consider the geodesic
whose path is the circle $|z-1+iR|^2 = R^2$ for $R = \cot(\pi/12) = 2
+ \sqrt{3}$ taken as ending at $z = \exp(-5\pi i/6)$ and starting at
$z=1$, points which can be easily identified on the disk on the right
in Figure~1.  This motion consists of exactly three Kasner epochs
delimited by two bounces against potential walls.  Likewise, from any
point in the fundamental domain there are countably many geodesics
that in the forward direction have only a finite number of Kasner
epochs.  These are the geodesics that end at points on the unit
circle where two images of the fundamental potential walls meet. Such
points are found by bisecting a finite number of times the arcs on
$|z|=1$ bounded by larger images of the fundamental potential walls.
[These trajectories correspond the those in the Belinskii,
Khalatnikov, and Lifshitz (1970) description with rational values of
$u$ in equations~(\ref{e-BKL}).]  For these trajectories the
hypothesis that only one potential wall at a time need be considered
is not convincingly justified since asymptotically the trajectories
follow a path between two walls.  Although the tangent directions
$(x,y,\theta)$ giving rise to such trajectories are a set of measure
zero (two dimensional in a three dimensional space) it is an
intriguing question whether any solution except Taub's has only a
finite number of oscillations approaching the singularity.

Let us study in more detail trajectories that have taken their last
bounce against a potential wall and are proceeding in the fundamental
domain on a geodesic ending at $z=1$. [These are just the circles
$|z-1 + iR|^2 = R^2$ for any real $R$ with $R^2 > 3$.]  To see what
has been neglected in this reduction to the Lobachevsky plane let us
try to find these trajectories in the description by
equation~(\ref{e-H}).  Translating $|z-1 + iR|^2 = R^2$ back to the
$\beta\Omega$ coordinates yields $\Omega-\beta_+ -R\beta_- = 0$ with
$R\beta_- > 0$.
Then set $p_\Omega = -1$ to normalize the independent
variable $\tau$. With the potential term neglected to give a geodesic
we can find $p_+$ and $p_-$ from ${\cal H} = 0$.  Two different null
geodesics in the minisuperspace metric $d\$^2$ project onto the same
geodesic in the Lobachevsky plane, a backward one leading out of the
fundamental region with $p_+ < 0$ and the one we want which is $p_+ =
1$, $p_- = 0$, so that $\beta_-$ remains constant as $\beta_+
\rightarrow +\infty$.  These considerations have neglected the
potential terms so that we have assumed $2{\cal H} = -p_\Omega^2 +
p_+^2 + p_-^2 $.  We will next reinstate some of the potential terms
neglected in the geodesic approximation.

As the geodesic of interest retreats from its last collision with a
potential wall, the motion is given by $\beta_+ = \tau$ and $\Omega =
\tau + R \beta_0$ with $\beta_- = \beta_0$ constant.
At $\tau = 0$ the condition $\Omega \gg 1$ assumed for the geodesic
approximation then gives $R\beta_0  \gg 1$.  In the
Hamiltonian~(\ref{e-H}) the potential term representing the last
potential wall encountered is $\frac{1}{6}\exp(-4\Omega - 8\beta_+) =
\frac{1}{6}\exp(-12\tau-4R\beta_0)$ which is readily neglected as
$\tau$ increases.  The potential walls between which the geodesic
leads are given by the term $\frac{1}{3}\exp(-4\Omega + 4\beta_+)
\cosh(4\sqrt{3}\beta_-) = \frac{1}{3} \exp(-4R\beta_0)
\cosh(4\sqrt{3}\beta_0)$ which is independent of $\tau$ in this
geodesic approximation, but not increasing in significance as would
be the case for a geodesic impinging on a potential wall.  One can
expect then that the neglect of this term is not justified for this
trajectory, since even a small force can have a large effect over a
long period of time, and the effects of the this term on the motion
are not projected to ever decrease while the motion is near the
geodesic.

To recognize that $-4\Omega + 4\beta_+$ is constant in the geodesic
approximation we introduce a new set of parameters $\xi\eta\beta_-$ to
replace the previous $\Omega\beta_+\beta_-$.  In
terms of these new parameters the Hamiltonian is
\begin{equation}\label{e-Hxi}
 {\cal H} = -p_\xi p_\eta + \half p_-^2 + {\cal V} \quad .
\end{equation}
with
\begin{eqnarray}\label{e-Vetaxi}
  {\cal V} & = &  \mbox{$\frac{1}{3}$}\exp(-4\sqrt{2}\eta)
                        [\cosh(4\sqrt{3}\beta_-)-1]
                  + \mbox{$\frac{1}{6}$}\exp[-2\sqrt{2}(3\xi - \eta)]
                    \nonumber\\
           &   &  - \mbox{$\frac{2}{3}$}\exp[-\sqrt{2}(3\xi + \eta)]
                    \cosh(2\sqrt{3}\beta_-)
                        \quad .
\end{eqnarray}
The canonical transformation that connects these two forms is
\begin{eqnarray}
     \xi = (\Omega + \beta_+)/\sqrt{2}\; , & \quad &
                    p_\xi = (p_\Omega + p_+)/\sqrt{2} \; ,\nonumber\\
     \eta =(\Omega - \beta_+)/\sqrt{2}\; , & \quad &
                    p_\eta =(p_\Omega - p_+)/\sqrt{2} \; .
\end{eqnarray}
In the geodesic approximation upon which we hope to improve these
variables have $\sqrt{2}\eta=R\beta_0$ and $\beta_-=\beta_0$ constant
while $\sqrt{2}\xi =2 \tau+R\beta_0$.  Thus the first term in ${\cal
V}$ from equation~(\ref{e-Vetaxi}) above is constant in this
approximation, the second term representing the prior potential wall
decreases as $\exp(-12\tau)$ and the third (negative) term which is
essential at the maximum of volume expansion decreases as
$\exp(-6\tau)$.  We will therefore neglect these last two terms and
study the motion with
\begin{equation}\label{e-Veta}
  {\cal V} =  \mbox{$\frac{1}{3}$}\exp(-4\sqrt{2}\eta)
                        [\cosh(4\sqrt{3}\beta_-)-1]
                        \quad .
\end{equation}
{}From the Hamiltonian~(\ref{e-Hxi}) with the approximate
potential~(\ref{e-Veta}) one sees that $-p_\xi = d\eta/d\tau$ is
constant even without the geodesic (${\cal V}=0$) approximation.
Also neither $p_\xi$ nor $-p_\eta = d\xi/d\tau$ can ever be zero
since ${\cal H}= 0$ and $\half p_-^2 + {\cal V}$ is positive (apart
from the Taub solutions with $p_- = 0 = \beta_-$).

We had chosen $\tau=0$ to be the stage $\beta_+ = 0$ of the
approximate geodesic motion where the previously important term in the
potential had decreased to the size of the term included in
equation~(\ref{e-Veta}).  To continue more accurately from this state
we choose as initial conditions at $\tau = 0$
$$
   \beta_- = \beta_0 \quad , \quad
   \sqrt{2}\eta = R\beta_0  =  \sqrt{2}\xi  \quad ,
$$
\begin{equation}
   p_- = 0 \quad , \quad -p_\eta = \sqrt{2} \quad , \quad
   -p_\xi \sqrt{2} = {\cal V}_0 \equiv \mbox{$\frac{1}{3}$}
            \exp(-4R\beta_0) [\cosh(4\sqrt{3}\beta_0)-1]
\end{equation}
retaining $-p_\eta = \sqrt{2}$ and $p_- = 0$ from the geodesic motion
so that only $-p_\xi$ is changed from zero to a small value which
will remain constant during the subsequent evolution.  (Recall from
the geodesic circles in $xy$ coordinates that $|R|>\sqrt{3}$ while
from our choice of $\tau = 0$ at $\beta_+ =0$ we had $R\beta_0 \gg
1$ so ${\cal V}_0 \ll 1$ follows.)

A central question in this study is whether motion under the
conditions of validity of the approximation~(\ref{e-Veta}) remain
valid for $\tau \rightarrow \infty$, or whether motion toward the
third potential wall at negative $\beta_+$ could evolve.  The
analysis requires the study of what I call the ``nonlinear Bessel
equation'' (NLB) which in zeroth order is
\begin{equation}\label{e-NLB}
   \frac{1}{r}\frac{d}{dr}\left(r\frac{db}{dr}\right) + \sinh b = 0
\end{equation}
and has, for small values of $b$, the solutions $b(r) = Z_0(r)$ where
$Z_0$ is a Bessel function as Chitre (1972b, see also Misner 1970)
noticed in studying this question.  To see how this equation arises
we must study the $\beta_-$ motion, but first we point out some
simpler aspects of this dynamics. In addition to the equation $-p_\xi
= $const\ \  that arises from the absence of $\xi$ from the
Hamiltonian~(\ref{e-Hxi}) in the approximation~(\ref{e-Veta}) one
finds also that
\begin{equation}
   \frac{d}{d\tau}(p_\xi p_\eta) = p_\xi \frac{dp_\eta}{d\tau}
   = - p_\xi \frac{\partial {\cal H}}{\partial\eta}
   = -4{\cal V}_0 {\cal V} < 0
\end{equation}
where from ${\cal H} = 0$ one has
\begin{equation}
   p_\eta p_\xi = \half p_-^2 + {\cal V} > 0 \quad .
\end{equation}
Thus $p_\eta p_\xi = -p_\eta {\cal V}_0/\sqrt{2}$ is a positive
quantity which monotonically decreases.  The NLB equation appears
when the Hamiltonian equations for $\beta_-$ are written.  They read
$d\beta_-/d\tau = p_-$ and $dp_-/d\tau = -\partial {\cal H}/\partial
\beta_-$ or
\begin{equation}
   \frac{d^2\beta_-}{d\tau^2} + \frac{4}{\sqrt{3}}\exp(-4\sqrt{2}\eta)
                                    \sinh(4\sqrt{3}\beta_-) = 0 \quad .
\end{equation}
Since $-p_\xi = d\eta/d\tau = {\cal V}_0 / \sqrt{2}$ we can, using
also the initial conditions assumed at $\tau = 0$ write
\begin{equation}
   \sqrt{2}\eta = {\cal V}_0 \tau + R\beta_0
\end{equation}
and eliminate $\eta$ from this differential equation:
\begin{equation}
   \frac{d^2\beta_-}{d\tau^2}
            + \frac{4}{\sqrt{3}}\exp(-4{\cal V}_0 \tau -4 R\beta_0)
                                \sinh(4\sqrt{3}\beta_-) = 0 \quad .
\end{equation}
Then a rescaling of both $\beta_-$ and $\tau$ to $b=4\sqrt{3}\beta_-$
and $\nu = 2{\cal V}_0 \tau + 2R\beta_0 + \ln({\cal V}_0/2)$ gives
\begin{equation}\label{e-bnu}
   \frac{d^2 b}{d\nu^2} + \exp(-2\nu) \sinh(b) = 0 \quad .
\end{equation}
The substitution $r = \exp(-\nu)$ then leads to
equation~(\ref{e-NLB}) which was first derived in this connection by
Belinskii and Khalatnikov (1969). Note that the change from $\tau$
to $\nu$ as a natural independent variable corresponds to a change in
time scale from $\Delta\tau \approx 1$ for the prior Kasner epochs to
$\Delta\nu \approx 1$ or $\Delta\tau \approx 1/2{\cal V}_0 \gg 1$ for
any subsequent oscillations in $\beta_-$.

Equation~(\ref{e-bnu}) for $\nu \ll -1$ or equation~(\ref{e-NLB}) for
$r \gg 1$ has a simple behavior of oscillations in an adiabatically
changing potential well where the oscillation period is short
compared to the time scale $\Delta\nu \approx 1$ on which the
potential changes.  This condition of negative initial $\nu$
arises for many but not all initial conditions, e.g., for $\beta_0^2
\ll 1$ or for $|R| > 2\sqrt{3}$.  Under these conditions the
amplitude $B$ of the $\beta_-$ oscillation increases with $B^2
e^{-\nu} \approx$ const\ \ in the (Bessel function) case of small
$B^2$ and with $B^2\exp(B-2\nu) \approx$ const\ \ for $B^2 \gg 1$
which is a long era of many Kasner epochs.  In all cases one expects
that for $\nu \gg 0$ the potential becomes negligible in
equation~(\ref{e-bnu}), and plausible asymptotic forms have been
given by Belinskii and Khalatnikov (1969), but it would be very
desirable to have rigorous mathematical results concerning the NLB
equation.  Solutions where $b$ remains bounded (like the
$J_0(e^{-\nu})$ Bessel function solution) can be found as power
series in $r=e^{-\nu}$:
\begin{equation}
   b = b_0 - \mbox{$\frac{1}{4}$} r^2 \sinh b_0 + 2^{-7} r^4 \sinh 2b_0
           + \mbox{$\frac{1}{3}$} 2^{-10 }r^6 (\mbox{$\frac{5}{3}$}
                \sinh b_0 - \sinh 3b_0 ) + O(r^8)
\end{equation}
for which a positive radius of convergence $r^2 < \half
\exp(-|b_0|)$ looks plausible.  These solutions become $y=0$ or $u=1$
geodesics in the Lobachevsky plane and return again after one bounce
against the potential wall at $x = -2+\sqrt{3}$ to the domain of
validity of equation~(\ref{e-bnu}) with revised initial conditions.
Solutions with $b \propto \nu$ are not known to me in any systematic
asymptotic series.  An approximation
\begin{equation}
   b =  2 k (\nu - \nu_o)
        - \frac{\exp[-2(1-|k|)\nu-2|k|\nu_o]}{8 (1-|k|)^2}
\end{equation}
gives a small value $-\exp[-4(1-|k|)\nu-4|k|\nu_o]/16(1-|k|)^2$ to
the left hand side of equation~(\ref{e-bnu}) when $\nu \gg 1$ and
$|k|<1$ and thus is a plausible asymptotic form. It corresponds to
the start of a new geodesic era.  If, as seems likely, every solution
of the NLB equation has one of these asymptotic forms for $\nu
\rightarrow \infty$ then, apart from the Taub solutions with $b=0$,
every solution returns to geodesic motion in the Lobachevsky plane as
the approximation~(\ref{e-Veta}) fails when $2\sqrt{2}(\eta - 3\xi)
=(1-k^2)\nu + O(1)$ becomes sufficiently large.

Let us summarize this analysis of the (two dimensional, zero measure)
set of tangent directions to the Lobachevsky plane that produce
geodesics which reach the singularity after only a finite number of
encounters with the potential walls.  For these cosmologies the
geodesic approximation (or the equivalent Kasner oscillations
generated by $u \mapsto u-1$ and $u \mapsto 1/u$) are not adequate.
The $u=\infty$ state does not persist when a better approximation is
employed, and in many cases a new regime of Kasner oscillations will
begin on a much longer time scale in $\tau$.  There is no plausible
suggestion that any solution except the Taub solutions can reach the
singularity while encountering the potential walls only a finite
number of times.  To show this more convincingly would require a
rigorous qualitative description of all solutions of the NLB
equation~(\ref{e-NLB}).

As a final comment I note that the measure~(\ref{e-measure}) which
allowed ergodic properties to be stated in the Lobachevsky plane
approximation arises out of the Einstein equation dynamics.  This
dynamics is expressed as a Hamiltonian for motion in minisuperspace
which is unique up to a conformal factor and displays a similarly
unique minisuperspace metric.  Approaching the singularity $t
\rightarrow \infty$ an asymptotic symmetry of $t$ independence for
the potential and conformal metric arises.  The minisuperspace metric
that displays this symmetry as a Killing vector is unique and defines
the measure we use in stating ergodic properties.  If a dynamically
defined measure on initial conditions near the singularity existed
more generally it would be very significant.  For example conjectures
in nonquantum relativity such as ``for almost all initial conditions
at the singularity inflationary evolution leads to regions of
spacetime approximating the spatially flat FRW cosmologies'' could
have a precise meaning.  Important questions on quantum geometry such
as factor ordering or the choice of an inner product might also
become more tractable if a natural measure on initial states were
available.  Thus further examples would be very useful where the
existence of asymptotic symmetries was decidable.

\end{document}